\documentstyle[epsf,epsfig]{article}

\newcommand{\bq}{\begin{equation}}
\newcommand{\eq}{\end{equation}}
\newcommand{\bqn}{\begin{eqnarray}}
\newcommand{\eqn}{\end{eqnarray}}

\newcommand{\lb}{\label}

         \def\gappr{\lower 3pt\hbox{$\buildrel > \over \sim\;$}}
         \def\gappl{\lower 3pt\hbox{$\buildrel < \over \sim\;$}}
         \def\gappeq{\mathrel{\rlap {\raise.5ex\hbox{$>$}} 
                               {\lower.5ex\hbox{$\sim$}}}}
         \def\lappeq{\mathrel{\rlap{\raise.5ex\hbox{$<$}}
                              {\lower.5ex\hbox{$\sim$}}}}
         \def\gsim{\lower.8ex\hbox{$\sim$}\kern-.75em\raise.45ex\hbox{$>$}\;}
         \def\lsim{\lower.8ex\hbox{$\sim$}\kern-.8em\raise.45ex\hbox{$<$}\;}
         \def\ltsima{$\; \buildrel < \over \sim \;$}
         \def\simlt{\lower.5ex\hbox{\ltsima}}
         \def\rtsima{$\; \buildrel > \over \sim \;$}
         \def\simrt{\lower.5ex\hbox{\rtsima}}
         \def\ln{{\rm ln}}
         \def\ex{{\rm e}}
         \def\dal{\hbox{$\sqcup$\hbox to 0pt{\hss$\sqcap$}}}

         \def\tanh{{\rm tanh}}
         
         \def\cosh{{\rm cosh}}
         
         \def\rlim{\lower 9pt\hbox{$\buildrel{\textstyle\longrightarrow}
                      \over{\scriptscriptstyle ~~r\rightarrow0~~}\;$}}
  
\begin{document}

\title{Gravitational Collapse of a
 Massless Scalar Field and a 
 Perfect Fluid with Self-Similarity of  
 the First Kind in (2+1) Dimensions}

\author{
F. I. M. Pereira$^1$ \thanks{E-mail Address: flavio@on.br} and\\
R. Chan$^1$ \thanks{E-mail Address: chan@on.br}\\
{\small $^1$ Coordena\c c\~ao de Astronomia e Astrof\'{\i}sica, Observat\'orio 
Nacional,} \\
\small Rua General Jos\'e Cristino 77, S\~ao Crist\'ov\~ao, CEP 20921--400, \\
\small Rio de Janeiro, RJ, Brazil}
\date{\today}
\maketitle

\begin{abstract}
Self-similar solutions of a collapsing perfect fluid and a massless scalar 
field  with kinematic self-similarity of the first kind in 2+1 dimensions are 
obtained. Their local and global properties of the solutions are studied. 
It is found that some of them represent gravitational collapse, in which black 
holes are always formed, and some may be interpreted as representing  cosmological models.
\end{abstract}
 
\section{Introduction}
 
One of the most interesting problems in gravitation theory is the
study of the relation that exists between the critical phenomena
and the process of black hole formation.
The studies of non-linearity of the Einstein field equations near
the threshold of
black hole formation reveal very rich phenomena \cite{Chop93a}-\cite{Chop93c},
which are quite similar to critical phenomena in Statistical
Mechanics and Quantum Field Theory \cite{Golden0}-\cite{Golden1}.
In particular, by numerically studying the gravitational collapse of a massless
scalar field in $3+1$-dimensional spherically symmetric
spacetimes, Choptuik found that the mass of such formed black
holes takes a scaling form,
$M_{BH} = C(p)\left(p -p^{*}\right)^{\gamma}$,
where $C(p)$ is a constant and depends on the initial data, and
$p$ parameterizes a family of initial data in such a way that when
$p > p^{*}$  black holes are formed, and when $p < p^{*}$ no black
holes are formed. It was shown that, in contrast to $C(p)$, the
exponent $\gamma$ is universal to all the families of initial
data studied. Numerically it was determined as $\gamma \sim 0.37$.
The solution with $p = p^{*}$, usually called the critical
solution, is found also universal. Moreover, for the massless
scalar field it is periodic, too. Universality  of the critical
solution and exponent, as well as the power-law scaling of the
black hole mass all have given rise to the name {\em Critical
Phenomena in Gravitational Collapse}. Choptuik's studies were soon
generalized to other matter fields \cite{Gun00,Wang01}, and now
the following seems clear: (a) There are two types of critical
collapse, depending on whether the black hole mass takes the
scaling form ($M_{BH}$) or not. When it takes the scaling form,
the corresponding collapse is called Type $II$ collapse, and when
it does not it is called Type $I$ collapse. In the type $II$
collapse, all the critical solutions found so far have either
discrete self-similarity (DSS) or homothetic self-similarity
(HSS), depending on the matter fields. In the type $I$ collapse,
the critical solutions have neither DSS nor HSS. For certain
matter fields, these two types of collapse can co-exist. (b) For
Type $II$ collapse, the corresponding exponent is universal only
with respect to certain matter fields. Usually, different matter
fields have different critical solutions and, in the sequel,
different exponents. But for a given matter field the critical
solution and the exponent are universal.  So far, the
studies have been mainly restricted to spherically symmetric case
and it is not clear whether or not the critical solution and exponent are
universal with respect to different symmetries of the
spacetimes \cite{Cho03a,Cho03b,Wang03}. (c)  A critical solution for both of 
the two
types has one and only one unstable mode. This now is considered as one
of the main criteria for a solution to be critical. (d) The
universality of the exponent is closely  related to the last
property. In fact, using dimensional analysis \cite{Even0}-\cite{Even3} one can
show that $\gamma = \frac{1}{\left|k\right|}$,
where $k$ denotes the unstable mode.

From the above, one can see that to study (Type $II$)
critical collapse, one
may first find some particular solutions by imposing certain
symmetries, such as, DSS or HSS. Usually this considerably
simplifies the problem. For example, in the spherically symmetric
case, by imposing HSS symmetry the Einstein field equations can be
reduced from PDE's to ODE's.    Once the particular solutions are
known, one can study their linear perturbations and find out the
spectrum of the corresponding eigenmodes. If a solution has one
and only one unstable mode, by definition we may consider it as a
critical solution (See also the discussions given in
\cite{Brady02}). The studies of critical collapse  have been
mainly numerical so far, and analytical ones are still highly
hindered by the complexity of the problem, even after imposing
some symmetries.

Lately, Pretorius and Choptuik (PC) \cite{PC00} studied
gravitational collapse of a massless scalar field in an anti-de
Sitter background in $2+1$-dimensional spacetimes with circular
symmetry, and found that the  collapse exhibits  critical
phenomena and the mass of such formed black holes takes the
scaling form ($M_{BH}$) with $\gamma = 1.2 \pm 0.02$, which
is different from that of the corresponding $3+1$-dimensional
case.  In addition, the critical solution is also different, and,
instead of having DSS, now has HSS. The above results were
confirmed by independent numerical studies \cite{HO01}. However,
the exponent obtained by Husain and Olivier (HO),  $\gamma \sim
0.81$,  is quite different from the one obtained by  PC. It is not
clear whether the difference is due to numerical errors or to some
unknown physics.

After the above numerical work, analytical studies of the same
problem soon followed up \cite{Gar01,CF01a,CF01b,CF01c,GG02,HWW04}. In
particular, Garfinkle found a class, say, $S[n]$, of exact
solutions to the Einstein-massless-scalar field equations and
showed that in the strong field regime the $n = 4$ solution fits
very well with the numerical critical solution found by PC.
Lately, Garfinkle and Gundlach (GG) studied their linear
perturbations and found that only the solution with $n = 2$ has
one unstable mode, while the one with $n = 4$ has three
\cite{GG02}. According to $\gamma = \frac{1}{\left|k\right|}$,
the corresponding
exponent is given by $\gamma = 1/|k| = 4/3$. Independently,
Hirschmann, Wang \& Wu (HWW) systematically
studied the problem, and  found that the $n = 4$ solution indeed
has only one unstable mode \cite{HWW04}. This difference actually
comes from the use of different boundary conditions. As a matter
of fact, in addition to the ones imposed by GG \cite{GG02}, HWW
further required that no matter field should come out of the
already formed black holes. This additional condition seems
physically quite reasonable and has been widely used in the
studies of black hole perturbations \cite{Chandra83}. However, now
the corresponding exponent is given by $\gamma = 1/|k| = 4$,
which is significantly different from the numerical ones. So far,
no  explanations about these differences have been worked out,
yet.

Self-similarity is usually
divided into two classes, one is the discrete self-similarity
mentioned above, and the other is the so-called kinematic
self-similarity (KSS) \cite{CH89a,CH89b}, and sometimes it is also called
continuous self-similarity (CSS). KSS or CSS is further classified
into three different kinds, the zeroth, first and second. The
kinematic self-similarity of the first kind is also called
homothetic self-similarity, first introduced to General Relativity
by Cahill and Taub in 1971 \cite{CT71}. In Statistical Mechanics,
critical solutions with KSS of the second kind seem more generic
than those of the first kind \cite{Golden0}-\cite{Golden1}. However,  critical
solutions with KSS of the second kind have not been found so far
in gravitational collapse, and it would be very interesting to
look for such solutions.   

Besides, recently we have studied gravitational collapse of perfect fluid
with kinematic self-similarities in four-dimensional spacetimes
\cite{Carlos02}, a subject that has been recently studied
intensively (for example, see  \cite{fluid0}-\cite{fluid3} and references
therein.). In this paper, we shall investigate the same problem
but in 2+1 gravity \cite{Carlip,Yasuda} 
and including a massless scalar field
\cite{Pereira} \cite{CFJW05}. The main motivation of such a
study comes from recent investigation of critical collapse of a
scalar field in 3D gravity \cite{PC00,HO01,Gar01,CF01a,CF01b,CF01c,GG02,HWW04}.

We shall present in this work the self-similar solutions of a collapsing 
perfect fluid and a massless scalar field  with kinematic self-similarity of 
the first kind in 2+1 dimensions.  Their local and global properties of the 
solutions are studied. Specifically, in section II we present the self-similar Einstein's field equations,
and in section III we show the general solutions obtained.  In section IV we present the global properties 
of the solutions and in Section V we conclude our work.

\section{Field Equations}

The 2+1-dimensional spacetimes with circular symmetry and zero-rotation are described by the metric
\begin{equation}
ds^2=\gamma_{ab}(t,r)dx^a dx^b+g_{\theta\theta}(t,r)d\theta^2~,
\label{metr1}
\end{equation}
where $\{x^a\}=\{t,r\}$, ($a,b=0, 1$), and $\theta$ denotes the angular coordinate, with the 
hypersurfaces $\theta=0,~2\pi$ being identified.  The metric is invariant under the following 
coordinate transformation,

\begin{equation}
t=t(t^\prime,r^\prime),~~~r=r(t^\prime,r^\prime)~.
\label{gf}
\end{equation}

On the other hand, the energy-momentum tensor for a perfect fluid is given by
\begin{equation}
T_{\mu\nu}(\rho,p)=(\rho+p)u_\mu u_\nu-pg_{\mu\nu}~,
\label{Tmn1}
\end{equation}
where $u_\mu$ is the velocity of the fluid, $\rho$ and $p$ are its energy density and pressure.
The energy-momentum tensor for a massless scalar field is given by

\begin{equation}
T_{\mu\nu}(\phi)=\phi_{,\mu}\phi_{,\nu}-\frac{1}{2}g_{\mu\nu}\phi_{,\alpha}\phi^{,\alpha}~,
\label{Tmn2}
\end{equation}
where $(~)_{,\alpha}=\partial(~)/\partial_\alpha$.  We choose the coordinates such that 

\begin{equation}
u_\mu\equiv \sqrt{g_{00}}~\delta^0_\mu~,~~{\rm and}~~g_{01}=0~,
\label{ug}
\end{equation}
which implies that the coordinates are comoving with the perfect fluid.
Then, the metric (\ref{metr1}) can be cast in the form
\begin{equation}
ds^2=l^2\bigg\{{\rm e}^{2\Phi(t,r)}dt^2-\ex^{2\Psi(t,r)}dr^2-r^2S^2(t,r)d\theta^2\bigg\}~,
\label{metr2}
\end{equation}
where $l$ is an unit constant with the dimension of length, so that all the coordinates 
$\{x^a\}=\{t,r,\theta\}$ are dimensionless.  We construct a projector operator 
$h_{\mu\nu}$ by

\begin{equation}
h_{\mu\nu}\equiv g_{\mu\nu}-u_\mu u_\nu~,
\label{h}
\end{equation}
from which we find $h^{\alpha\beta}u_\alpha u_\beta=0$.  Once defined the projection 
operator $h_{\mu\nu}$, following Carter and Henriksen \cite{CH89a,CH89b}, we define the kinematic 
self-similarity by

\begin{equation}
{\cal L}_\xi h_{\mu\nu}=2h_{\mu\nu},~~~~{\cal L}_\xi u^\mu=-\alpha u^\mu~,
\label{lie}
\end{equation}
where ${\cal L}_\xi$ denotes the Lie differentiation along the vector $\xi^\mu$, $\alpha$ 
is a {\it dimensionless} constant.  When $\alpha=0$, the self-similarity is said to be of 
the {\it zeroth kind}, when $\alpha=1$ it is of the {\it first kind (or homothetic)}, and 
otherwise it is of the {\it second kind}.  In this work we shall consider only solutions 
with self-similarity of the first kind ($\alpha=1$).  Applying the above definition 
to the metric (\ref{metr2}), we find that 

\begin{equation}
\Phi(t,r)=\Phi(x)~,~~~~\Psi(t,r)=\Psi(x)~,~~~~S(t,r)=S(x)~,
\label{fps2}
\end{equation}
where the self-similar variable $x$ and $\xi^\mu$ are given by

\begin{equation}
x=\ln\bigg(\frac{r}{-t}\bigg)~,
\label{x}
\end{equation}
and
\begin{equation}
\xi^\mu\frac{\partial}{\partial x^\mu}= t\frac{\partial}{\partial t}+
r\frac{\partial}{\partial r}~.
\label{kill}
\end{equation}

Before looking for solutions of the Einstein field equations, we must take into account 
that for the metric to represent circular symmetry some physical and geometrical conditions 
must be imposed \cite{Fatima0}-\cite{Fatima9}.  We impose the following regularity conditions 
for the gravitational collapse:

(i) There must exist a symmetry axis, which can be expressed as

\begin{equation}
{\cal R}\equiv|\xi^\mu_{(\theta)}\xi^\nu_{(\theta)}g_{\mu\nu}|^{1/2}\rightarrow 0,
\label{Rxixi}
\end{equation}
as $r\rightarrow0$, where we have chosen the radial coordinate such that the axis is located 
at $r=0$, and $\xi^\mu_{(\theta)}$ is the Killing vector with closed orbit given by 
$\xi^\alpha_{(\theta)}\partial_\alpha=\partial_\theta$.

(ii) The spacetime near the symmetry axis is locally flat, which can be written as \cite{Kramer80}

\begin{equation}
{\cal R}_{,\alpha}{\cal R}_{,\beta}g^{\alpha\beta}\rightarrow-1~,
\label{RRg}
\end{equation}
as $r\rightarrow0$.  Note that solutions failing to satisfy this condition sometimes are 
also acceptable.  For example, when the right-hand side of the above equation approaches 
a finite constant, the singularity at $r=0$ may be related to a point-like particle \cite{VS}.
However, since here we are mainly interested in gravitational collapse, in this paper we 
shall assume that this condition holds strictly at the beginning of the collapse, so that 
we can be sure that the singularity to be formed later on the axis is due to the collapse.

(iii) No closed timelike curves.  In spacetimes with circular symmetry, closed timelike 
curves can be easily introduced.  To ensure their absence, we assume that the condition  

\begin{equation}
\xi^\mu_{(\theta)}\xi^\nu_{(\theta)}g_{\mu\nu}<0~,
\label{CTC}
\end{equation}
holds in the whole spacetime.

In addition to these conditions, it is usually also required that the spacetime be 
asymptotically flat in the radial direction.  However, since we consider solutions with 
self-similarity, this condition cannot be satisfied, unless we restrict their validity 
only up to a maximal radius, say, $r=r_0(t)$, and then join them with others in the region 
$r>r_0(t)$, which are asymptotically flat as $r\rightarrow\infty$.  In this paper, we 
shall not consider such a possibility, and simply assume that the self-similar solutions 
are valid in the whole spacetime.

\section{Self-Similar Solutions of the First Kind}

In this section, we study solutions with self-similarity of the first kind.  To 
obtain the desired equations we substitute equations (\ref{Tmn1}), (\ref{Tmn2}) and (\ref{ug}) 
into the Einstein field equations
\begin{equation}
G_{\mu\nu}=\kappa[T_{\mu\nu}(\rho,p)+T_{\mu\nu}(\phi)],
\end{equation}
and also using the Klein-Gordon equation
\begin{equation}
\dal\phi=0~,
\label{Gmn}
\end{equation}
where $\kappa$ is the Einstein coupling constant,
$\dal=g^{\alpha\beta}\nabla_\alpha\nabla_\beta$ with $\nabla_\alpha$ being the 
covariant derivative.  When $\alpha=1$, according to equation (\ref{x}) terms with 
powers of $r$  can be substituted by the same powers of $(-t)$ since $r=\ex^x(-t)$.  
Then it can be shown that the Einstein field equations in this case become (see Appendix A)

\bqn
\lb{eins1}
\ex^{2(x-\Phi)}\big[y\Psi_{,x}-\frac{\kappa}{2}(\phi_{,x}+\phi_{,\tau})^2\big]
-\ex^{-2\Psi}\big[y_{,x}+(1+y)(y-\Psi_{,x})+\frac{\kappa}{2}(\phi_{,x})^2\big]
&=&\frac{\kappa}{2}~\rho_0~,\\
\lb{eins2}
-\ex^{2(x-\Phi)}\big[y_{,x}+y(y-\Phi_{,x}+1)+\frac{\kappa}{2}(\phi_{,x}+\phi_{,\tau})^2\big]
+\ex^{-2\Psi}\big[(1+y)\Phi_{,x}-\frac{\kappa}{2}(\phi_{,x})^2\big]&=&\frac{\kappa}{2}~p_0~,\\
\lb{eins3}
-\ex^{2(x-\Phi)}\big[\Psi_{,xx}+\Psi_{,x}(\Psi_{,x}-\Phi_{,x}+1)+\frac{\kappa}{2}(\phi_{,x}+\phi_{,\tau})^2\big]&+&\nonumber\\
\ex^{-2\Psi}\big[\Phi_{,xx}+\Phi_{,x}(\Phi_{,x}-\Psi_{,x}-1)+\frac{\kappa}{2}(\phi_{,x})^2\big]
&=&\frac{\kappa}{2}~p_0~,\\
\lb{eins4}
y_{,x}+(1+y)(y-\Psi_{,x})-y\Phi_{,x}+\kappa~\phi_{,x}(\phi_{,x}+\phi_{,\tau})&=&0~,\\
\lb{eins5}
\ex^{2(x-\Phi)}\big[(\phi_{,x}+\phi_{,\tau})_{,x}-(\Phi_{,x}-\Psi_{,x}-y-1)(\phi_{,x}+
\phi_{,\tau})\big]&-&\nonumber\\
\ex^{-2\Psi}\big[\phi_{,xx}+(\Phi_{,x}-\Psi_{,x}+y)\phi_{,x}\big]&=&0~.
\eqn

In order to determinate the metric coefficients and the $x$ dependence 
of the scalar field completely, and to make possible the calculations, 
we assume the equation of state for the perfect fluid given by

\bq
p=\omega\rho~,
\lb{eos}
\eq
for any $\omega$.

We use for the massless scalar field the general form $\phi(\tau,x)=a\tau+F(x)$,  where $a$=constant.

To solve the above equations in the general case is not an easy task. In the following, we shall 
consider several particular cases. We first show some particular solutions for a dust fluid and 
later for the general case with $\omega\neq0$. Solutions corresponding to $p=\rho=0$, that have already 
been considered in reference \cite{HWW04}, are also shown in section 3.3.   

\subsection{Dust Fluid ($p=0$)}

From geodesic equations we find that $\Phi(x)=\Phi_0$, constant.  Then, 
the Einstein equations are given by

\bqn
\lb{eins1p}
\ex^{2(x-\Phi_0)}\big[y\Psi_{,x}-\frac{\kappa}{2}(\phi_{,x}+\phi_{,\tau})^2\big]
-\ex^{-2\Psi}\big[y_{,x}+(1+y)(y-\Psi_{,x})+\frac{\kappa}{2}(\phi_{,x})^2\big]
&=&\frac{\kappa}{2}~\rho_0~,\\
\lb{eins2p}
-\ex^{2(x-\Phi_0)}\big[y_{,x}+y(y+1)+\frac{\kappa}{2}(\phi_{,x}+\phi_{,\tau})^2\big]
-\ex^{-2\Psi}\frac{\kappa}{2}(\phi_{,x})^2&=&0~,\\
\lb{eins3p}
-\ex^{2(x-\Phi_0)}\big[\Psi_{,xx}+\Psi_{,x}(\Psi_{,x}+1)+\frac{\kappa}{2}(\phi_{,x}+\phi_{,\tau})^2\big]
+\ex^{-2\Psi}\frac{\kappa}{2}(\phi_{,x})^2&=&0~,\\
\lb{eins4p}
y_{,x}+(1+y)(y-\Psi_{,x})+\kappa~\phi_{,x}(\phi_{,x}+\phi_{,\tau})&=&0~,\\
\lb{eins5p}
\ex^{2(x-\Phi_0)}\big[(\phi_{,x}+\phi_{,\tau})_{,x}+(\Psi_{,x}+y+1)(\phi_{,x}+\phi_{,\tau})\big]
-\ex^{-2\Psi}\big[\phi_{,xx}-(\Psi_{,x}-y)\phi_{,x}\big]&=&0~.
\eqn

In the following, we consider several particular cases.

\subsubsection{Solution 1}

\begin{eqnarray}
\label{Ip}
\Phi(x)&=&\Phi_0~\nonumber\\
\Psi(x)&=&-x+\Psi_0~\nonumber\\
y(x)&=&-1~\nonumber\\
S(x)&=&S_0\ex^{-x}~\\
\phi(x,\tau)&=&\phi_0\nonumber\\
\rho&=&\frac{\ex^{-2\Phi_0}}{\kappa~l^2(-t)^2}~\nonumber\\
p&=&0~\nonumber
\end{eqnarray}
where $\Phi_0$, $\Psi_0$, $S_0$ and $\phi_0$ are arbitrary integration constants.

\subsubsection{Solution 2}

\begin{eqnarray}
\label{IIIp}
\Phi(x)&=&\Phi_0~\nonumber\\
\Psi(x)&=&\ln\big[\cosh(\frac{x-x_0}{2})\big]-\frac{x}{2}+\Psi_0~\nonumber\\
y(x)&=&-1~\nonumber\\
S(x)&=&S_0\ex^{-x}~\\
\phi(x,\tau)&=&\phi_0\nonumber\\
\rho&=&\frac{\ex^{-2\Phi_0}}{2~\kappa~l^2(-t)^2}
\big[1-\tanh(\frac{x-x_0}{2})\big]~\nonumber\\
p&=&~0~\nonumber
\end{eqnarray}

\subsubsection{Solution 3}

\begin{eqnarray}
\label{Vp}
\Phi(x)&=&\Phi_0~\nonumber\\
\Psi(x)&=&-a\sqrt{\frac{\kappa}{2}}x+\Psi_0~\nonumber\\
y(x)&=&a\sqrt{\frac{\kappa}{2}}-1~\nonumber\\
S(x)&=&S_0\ex^{y~x}~\\
\phi(x,\tau)&=&a\tau+\phi_0~\nonumber\\
\rho&=&0~\nonumber\\
p&=&0,~\nonumber \\
a&=&\frac{1}{\sqrt{2K}} \nonumber
\end{eqnarray}

\subsection{$p=\omega\rho$}

 Now, let us consider particular solutions for $\omega\neq0$. Thus, we will
use the general equations (\ref{eins1})-(\ref{eins5}).

\subsubsection{Solution 4}

\begin{eqnarray}
\label{I}
\Phi(x)&=&\Phi_0~\nonumber\\
\Psi(x)&=&\frac{-x}{\omega+1}+\Psi_0~\nonumber\\
y(x)&=&\frac{-1}{\omega+1}~\nonumber\\
S(x)&=&S_0\ex^{yx}~\\
\phi(x,\tau)&=&\phi_0\nonumber\\
\rho&=&\frac{\ex^{-2\Phi_0}}{\kappa~l^2(\omega+1)^2(-t)^2}~\nonumber\\
p&=&\omega\rho~\nonumber
\end{eqnarray}

\subsubsection{Solution 5}

\begin{eqnarray}
\label{II}
\Phi(x)&=&\Phi_0~\nonumber\\
\Psi(x)&=&-\frac{x}{2}+\Psi_0~\nonumber\\
y(x)&=&-\frac{1}{2}~\nonumber\\
S(x)&=&S_0\ex^{-x/2}~\\
\phi(x,\tau)&=&a~\tau+\phi_0\nonumber\\
\rho&=&\frac{\ex^{-2\Phi_0}}{4\kappa~l^2(-t)^2}(1-2\kappa a^2)~\nonumber\\
p&=&\rho~\nonumber
\end{eqnarray}

\subsubsection{Solution 6}

\begin{eqnarray}
\label{III}
\Phi(x)&=&\pm\Phi_1\ln|\ex^{x+2\Psi_0}-1|+\kappa~a^2 x+\Phi_0~\nonumber\\
\Psi(x)&=&\Phi(x)-\frac{x}{2}+\Psi_0~\nonumber\\
y(x)&=&-1/2~\nonumber\\
S(x)&=&S_0\ex^{-x/2}~\\
\phi(x,\tau)&=&a(\tau-x)+\phi_0\nonumber\\
\rho&=&p~=~-\frac{\ex^{-2\Phi_0}(2\kappa~a^2-1\pm2\Phi_1)}
{4~\kappa~l^2(-t)^{2(1-\kappa a^2)}~r^{2\kappa a^2}|\ex^{x+2\Psi_0}-1|^{\pm2\Phi_1}}\nonumber
\end{eqnarray}
with + if $x+2\Psi_0>0$ and - if $x+2\Psi_0<0$

\subsubsection{Solution 7}

\begin{eqnarray}
\label{IV}
\Phi(x)&=&\Phi_0~\nonumber\\
\Psi(x)&=&-x+\Psi_0~,~~~~\Psi_0=\Phi_0~\nonumber\\
y(x)&=&\frac{-1}{1+\kappa~a^2}\nonumber\\
S(x)&=&S_0\ex^{yx}~\nonumber\\
\phi(x,\tau)&=&a(\tau+yx)+\phi_0\\
\rho&=&\frac{\ex^{-2\Phi_0}}{\kappa~l^2(-t)^2}(1-y)\frac{1+2y}{2y}\nonumber\\
p&=&\omega\rho\nonumber\\
\omega&=&\frac{1+y}{1-y}\nonumber\\
a&=&\pm\sqrt{-\frac{1+y}{\kappa~y}}~,~~~~-1\leq y<0\nonumber
\end{eqnarray}

\bigskip

\subsection{$p=\rho=0$}

\subsubsection{Solution 8}

\begin{eqnarray}
\label{IVp}
\Phi(x)&=&\Phi_0~\nonumber\\
\Psi(x)&=&\Psi_0~\nonumber\\
y(x)&=&\frac{1}{2}[\tanh(\frac{x-x_0}{2})-1]~\nonumber\\
S(x)&=&S_0\ex^{-x/2}\cosh(\frac{x-x_0}{2})~\\
\phi(x,\tau)&=&\phi_0~\nonumber\\
\rho&=&p~=~0~\nonumber
\end{eqnarray}

\subsubsection{Solution 9}

\begin{eqnarray}
\label{IIp}
\Phi(x)&=&\Phi_0~\nonumber\\
\Psi(x)&=&-\frac{x}{2}+\Psi_0~\nonumber\\
y(x)&=&-1/2~\nonumber\\
S(x)&=&S_0\ex^{-x/2}~\\
\phi(x,\tau)&=&a\tau+\phi_0~\nonumber\\
\rho&=&p=~0~\nonumber\\
a&=&\pm \frac{1}{\sqrt{\kappa}} \nonumber
\end{eqnarray}

\subsubsection{Solution 10}

\begin{eqnarray}
\label{VIp}
\Phi(x)&=&\Phi_0~\nonumber\\
\Psi(x)&=&-x+\Psi_0~,~~\Psi_0=\Phi_0~\nonumber\\
y(x)&=&-1/2~\nonumber\\
S(x)&=&S_0\ex^{-x/2}~\\
\phi(x,\tau)&=&a(\tau-x/2)+\phi_0~\nonumber\\
\rho&=&p~=~0~\nonumber \\
a&=&\pm \frac{1}{\sqrt{\kappa}} \nonumber
\end{eqnarray}

\subsubsection{Solution 11}

\begin{eqnarray}
\label{VIIp}
\Phi(x)&=&\frac{1-2\kappa a^2}{2}~\ln|\ex^{x+2\Psi_0}-1|+
\kappa a^2(x+2\Psi_0)+\Phi_0~\nonumber\\
\Psi(x)&=&\Phi(x)-\frac{x}{2}+\Psi_0~\nonumber\\
y(x)&=&-1/2~\nonumber\\
S(x)&=&S_0\ex^{-x/2}~\\
\phi(x,\tau)&=&a(\tau-x)+\phi_0~\nonumber\\
\rho&=&p~=~0~\nonumber
\end{eqnarray}

\subsubsection{Solution 12}

\begin{eqnarray}
\label{VIIIp}
\Phi(x)&=&\frac{x}{2}+\Phi_0~\nonumber\\
\Psi(x)&=&-\frac{x}{2}+\Psi_0~,~~\Psi_0=\Phi_0~\nonumber\\
y(x)&=&-1/2~\nonumber\\
S(x)&=&S_0\ex^{-x/2}~\\
\phi(x,\tau)&=&a(\tau-x/2)+\phi_0~\nonumber\\
\rho&=&p~=~0~\nonumber \\
a&=&\pm \frac{1}{\sqrt{\kappa}} \nonumber
\end{eqnarray}

\bigskip

\section{Properties of the Self-Similar Solutions}

In this section we shall analyze the global structure for the solutions 
found in the last section.

\subsection{Dust Fluid}

Let us first consider the solutions for $p=0$.

\subsubsection{Solution 1}

From equation (\ref{Ip}), we find that the corresponding metric can be written as

\bq
ds^2=l^2\bigg\{\ex^{2\Phi_0}dt^2-\frac{(-t)^2}{r^2}\bigg[\ex^{2\Psi_0}dr^2+S^2_0r^2d\theta^2\bigg]\bigg\}~.
\lb{metrIp}
\eq
The geometrical radius is given by

\bq
{\cal R}=lS_0r\ex^{-x}=lS_0(-t)~.
\lb{RIp}
\eq
Without loss of generality, we assume that $S_0>0$. Since ${\cal R}$ is a function of $t$ only, it 
is easy to see that ${\cal R}_{,t}$ is always positive,
since $-t > 0$.  The whole spacetime is trapped, as one 
can see  from the outgoing and ingoing null geodesics expansions, which are now given 
by the equations (see Appendix A)

\bq
\theta_l=-~\frac{\ex^{-\Phi_0}}{2g(-t)}~~~{\rm and}~~~
\theta_n=-~\frac{\ex^{-\Phi_0}}{2f(-t)}~,
\lb{outingIp}
\eq
from which we have that $\theta_l\theta_n~>~0$~.

From equation (\ref{Ip}) we see that the space-time is always singular when $(-t)\rightarrow0$. On the other 
hand, the following expression,

\bq
R_{,\alpha}R_{,\beta}~g^{\alpha\beta}=l^2S_0^2\ex^{-2x}\bigg[\ex^{-2\Phi_0}\frac{r^2}{(-t)^2}-
0\bigg]=l^2S_0^2\ex^{-2\Phi_0}~
\lb{regIp}
\eq
is always nonzero. In this solution we do not have an apparent horizon but a singularity at $t=0$.  
Thus, it may be interpreted as representing a cosmological model.

\subsubsection{Solution 2}

In the case of equation (\ref{IIIp}), the metric reads

\bq
ds^2=l^2\bigg\{\ex^{2\Phi_0}dt^2-\frac{\ex^{2\Psi_0+x_0}}{4}\bigg[\ex^{-x_0}+\frac{(-t)}{r}\bigg]^2dr^2
-S^2_0(-t)^2d\theta^2\bigg\}~.
\lb{metrIIIp}
\eq

The outgoing and ingoing null geodesics expansions

\bq
\theta_l=-~\frac{\ex^{-\Phi_0}}{2g(-t)}~~~{\rm and}~~~
\theta_n=-~\frac{\ex^{-\Phi_0}}{2f(-t)}
\lb{outingIIIp}
\eq
are such that $\theta_l\theta_n~>~0$~. Thus, there is no apparent horizon. 

 The geometrical radius is given by 

\bq 
R=lS_0r\ex^{-x}=lS_0(-t)~,
\lb{RIIIp}
\eq
where it is assumed that $S_0>0$. On the other hand, we see that the right hand side of the 
expression

\bq
R_{,\alpha}R_{,\beta}~g^{\alpha\beta}=l^2S_0^2\ex^{-2x}\bigg[\ex^{-2\Phi_0}\frac{r^2}{(-t)^2}-
0\bigg]=l^2S_0^2\ex^{-2\Phi_0}~,
\lb{regIIIp}
\eq
is always nonzero.

This case is similar to that of Solution 1. Thus, this solution, as in Solution 1, may be interpreted as representing a cosmological model.

\subsubsection{Solution 3}

We can see that this solution 
 is equal to Solution 9, which is a solution studied by Hirschmann, Wang \& Wu
\cite{HWW04}.
 
\subsection{$p=\omega\rho$}

Now, we consider the solutions of equations (\ref{I})-(\ref{IV}) 
for $w\neq0$.

\subsubsection{Solution 4}

From equation (\ref{I}) we see that for $0\leq\omega\leq1$ the space-time is the same as that obtained 
by Miguelote et al. \cite{Yasuda}, but for a constant scalar field which, 
without loss of generality, can be $\phi_0=0$. 

\subsubsection{Solution 5}

From equation (\ref{II}), we find that the corresponding metric can be written 
in the form

\bq
ds^2=l^2\bigg\{\ex^{2\Phi_0}dt^2-\frac{(-t)}{r}(\ex^{2\Psi_0}dr^2
+r^2S^2_0d\theta^2)\bigg\}~.
\lb{metrII}
\eq
The geometrical radius is given by

\bq
{\cal R}=lrS_0\ex^{-x/2}=lS_0r^{1/2}(-t)^{1/2}~.
\lb{RII}
\eq
We assume that $S_0>0$, since ${\cal R} > 0$.

Let us now investigate the behavior of the ingoing and outgoing expansions of 
the null geodesics. From equations (\ref{II}), (\ref{B2}) and (\ref{B3}) we find 
that

\begin{eqnarray}
\label{outII}
\theta_l=\frac{\ex^{-\Phi_0}}{4g(-t)}
\bigg[\bigg(\frac{r_{AH}}{r}\bigg)^{1/2}-1 \bigg]
\left\{
\begin{array}{l}
>0{~~\rm if}~~r<r_{AH}\\
=0{~~\rm if}~~r=r_{AH}~~~~~~\\
<0{~~\rm if}~~r>r_{AH}\\
\end{array}
\right.\
\end{eqnarray}
and
\begin{eqnarray}
\label{ingII}
\theta_n=-\frac{\ex^{-\Phi_0}}{4f(-t)}
\bigg[\bigg(\frac{r_{AH}}{r}\bigg)^{1/2}+1 \bigg]<0~,~{\rm for~any~}~r\in[0,~\infty)~.
\end{eqnarray}
where
\bq
r_{AH}=\ex^{2(\Phi_0-\Psi_0)}(-t)
\lb{AHII}
\eq
is the apparent horizon.  From the above equations we see that 
\begin{eqnarray}
\label{inoutIVa}
\theta_l\theta_n 
\left\{
\begin{array}{l}
>0{~~\rm if}~~r>r_{AH}~-~{\rm trapped}\\
=0{~~\rm if}~~r=r_{AH}~-~{\rm marginally~trapped}\\
<0{~~\rm if}~~r<r_{AH}~-~{\rm untrapped}\\
\end{array}
\right.\
\end{eqnarray}

In order to satisfy the regularity condition (\ref{RRg}) we find that
\begin{equation}
S_0=2\ex^{\Psi_0}~.
\label{RRGII}
\end{equation}

Thus, we conclude that we have a black hole.

\subsubsection{Solution 6} 

Due to the discontinuities in $\rho$ and $\theta_n$ at the horizon, this
solution does not represent a physical system.
 
\subsubsection{Solution 7}
  From equation (\ref{IV}) we find that the metric is given by
\bq
ds^2=l^2\bigg\{\ex^{2\Phi_0}~dt^2-\frac{(-t)^2}{r^2}\ex^{2\Psi_0}
~dr^2-S_0^2\frac{r^{2(1+y)}}{(-t)^{2y}}~d\theta^2\bigg\}~.
\lb{metrIV}
\eq

The geometrical radius is given by

\bq
{\cal R}=lrS_0\ex^{yx}=lS_0\frac{r^{1+y}}{(-t)^y}~,
\lb{RIV}
\eq
where we can see that ${\cal R}\rightarrow0$ when $r\rightarrow0$, since $-1<y<0$ because $a$ is real. We also assume $S_0 >0$.

  The outgoing and ingoing null geodesics expansions are given by

\begin{eqnarray}
\label{outIV}
\theta_l=\frac{\ex^{-\Psi(x)}}{2gr}\bigg(\frac{3\omega-1}{\omega+1}\bigg)
\left\{
\begin{array}{l}
>0{~~\rm if}~~\omega>1/3~~(y>-1/2)\\
=0{~~\rm if}~~\omega=1/3~~(y=-1/2)\\
<0{~~\rm if}~~\omega<1/3~~(y<-1/2)\\
\end{array}
\right.\
\end{eqnarray}
and
\begin{eqnarray}
\label{ingIV}
\theta_n=-\frac{\ex^{-\Psi(x)}}{2fr}~<0~~{\rm for~any}~~r\in[0,~\infty)~.
\end{eqnarray}
Thus, this solution does not present an apparent horizon and it represents
a cosmological solution.  

\subsubsection{Solutions 8}

As can be seen in equation (\ref{IVp}), this solution, corresponds to a flat 
space-time.  

\subsubsection{Solutions 9, 10, 11 e 12}

Since these solutions have already been studied by Hirschmann, Wang \& Wu 
\cite{HWW04}, we do not present any analysis of them.

\section{Conclusions}

In this paper we have obtained self-similar solutions of the Einstein field 
equations for a collapsing
massless scalar field and perfect fluid with kinematic self-similarity of the
first kind in 2+1
dimensions. The local and global properties of the solutions are studied.
Since in the general case it is not an easy task to solve the system of field
equations, we have considered
some particular cases.  We have found 12 solutions.  One of them does not
represent a physical system (Solution 6) and another one of them represents
a Minkowski spacetime (Solution 8). Six of them represent solutions 
already studied by others authors (Solutions 3, 4, 9, 10, 11 and 12).
Thus, we have obtained three new cosmological solutions (Solutions 1, 2 and 7)
and a new black hole solution (Solution 5).

\section*{Acknowledgments}

We would like to thank Dr. Anzhong Wang for the useful and helpful
discussions about this work.
One of the authors (RC) acknowledges the financial
support from FAPERJ (no. E-26/171.754/2000 and E-26/171.533/2002) and from
Conselho Nacional de Desenvolvimento Cient\'{\i}fico e Tecnol\'ogico - CNPq -
Brazil.
\bigskip

\appendix{\bf Appendix A - The field equations}

The non-zero components of the Einstein tensor are given by

\begin{eqnarray}
\label{Gtt}
G_{tt}&=&\frac{1}{(-t)^2}y\Psi_{,x}-
\frac{\ex^{2(\Phi-\Psi)}}{r^2}\big[y_{,x}+(1+y)(y-\Psi_{,x})\big]\\
\label{Grr}
G_{rr}&=&\frac{-\ex^{2(\Psi-\Phi)}}{(-t)^2}\big[y_{,x}+y(y-\Phi_{,x}+1)\big]+
\frac{1}{r^2}(1+y)\Phi_{,x}\\
\label{Gteta}
G_{\theta\theta}&=&S^2r^2\bigg\{-\frac{\ex^{-2\Phi}}{(-t)^2}\big[\Psi_{,xx}+
\Psi_{,x}(\Psi_{,x}-\Phi_{,x}+1)\big]\nonumber\\
&+&\frac{\ex^{-2\Psi}}{r^2}\big[\Phi_{,xx}+\Phi_{,x}(\Phi_{,x}-\Psi_{,x}-1)\big]
\bigg\}\\
\label{Gtr}
G_{tr}&=&~\frac{-1}{(-t)~r}\big[y_{,x}+(1+y)(y-\Psi_{,x})-y\Phi_{,x}\big]~,
\end{eqnarray}
where

\begin{equation}
y\equiv \frac{S_{,x}}{S}=(\ln S)_{,x}~,
\label{y}
\end{equation}
and the Klein-Gordon equation can be written as

\begin{eqnarray}
\label{KG}
\frac{\ex^{-2\Phi}}{(-t)^2}\big[\phi_{,xx}
-(\Phi_{,x}-\Psi_{,x}-y-1)(\phi_{,x}+\phi_{,\tau})\big]\nonumber\\
-\frac{\ex^{-2\Psi}}{r^2}\big[\phi_{,xx}+(\Phi_{,x}-\Psi_{,x}+y)\phi_{,x}\big]=0~.
\end{eqnarray}
The non-zero components of the energy-momentum tensor for a massless scalar 
field and a perfect fluid are given by

\begin{eqnarray}
\label{Ttt}
T_{tt}&=&\frac{1}{2~(-t)^2}~(\phi_{,x}+\phi_{,\tau})^2+
\frac{\ex^{2(\Phi-\Psi)}}{2~r^2}~(\phi_{,x})^2+\rho~\ex^{2\Phi}\\
\label{Trr}
T_{rr}&=&\frac{\ex^{2(\Psi-\Phi)}}{2~(-t)^2}~(\phi_{,x}+
\phi_{,\tau})^2+\frac{1}{2~r^2}~(\phi_{,x})^2+p~\ex^{2\Psi}\\
\label{Tteta}
T_{\theta\theta}&=&~r^2~S^2\bigg[\frac{\ex^{-2\Phi}}{2~(-t)^2}~(\phi_{,x}+
\phi_{,\tau})^2-\frac{\ex^{-2\Psi}}{2~r^2}~(\phi_{,x})^2+p~\bigg]\\
\label{Ttr}
T_{tr}&=&\frac{1}{r~(-t)}\phi_{,x}(\phi_{,x}+\phi_{,\tau})~.
\end{eqnarray}

In order to put the field equations in a more suitable form for calculations
we write the energy density and pressure as

\begin{eqnarray}
\label{prho}
\rho\equiv\frac{\rho_0(x)}{2~r^2}
~~~~~{\rm and}~~~~~p\equiv\frac{p_0(x)}{2~r^2}~.
\end{eqnarray}

For the self-similar solutions of the first kind the outgoing and ingoing expansion of 
the null geodesics and ${\cal R}_{,\alpha}{\cal R}_{,\beta}g^{\alpha\beta}$ 
are, respectively, given by \cite{Yasuda}

\begin{equation}
\theta_{l}=\frac{1}{2rg}\bigg\{(1+y)\ex^{-\Psi(x)}+
y\ex^{x-\Phi(x)}\bigg\}~,
\label{B2}
\end{equation}

\begin{equation}
\theta_{n}=\frac{-1}{2rf}\bigg\{(1+y)\ex^{-\Psi(x)}-
y\ex^{x-\Phi(x)}\bigg\}~,
\label{B3}
\end{equation}
where $f$ and $g$ are assumed to be positive functions, $f>0$ and $g>0$ 
(for a more detailed discussion, see appendix B of reference \cite{Yasuda}), 
and 

\begin{equation}
{\cal R}_{,\alpha}{\cal R}_{,\beta}g^{\alpha\beta}=S^2(x)\bigg\{
y^2\frac{r^2\ex^{-2\Phi(x)}}{(-t)^2}-
(1+y)^2\ex^{-2\Psi(x)}
\bigg\}~.
\label{B4}
\end{equation}


\begin{thebibliography}{99}
 
\bibitem{Chop93a} M. W. Choptuik, Phys. Rev. Lett. {\bf 70}, 9 (1993).

\bibitem{Chop93b} M. W. Choptuik, 
  ``{\em Critical Behavior in Massless
Scalar Field Collapse}," in {\it Approaches to Numerical
Relativity, Proceedings of the International Workshop on Numerical
Relativity}, Southampton, December, 1991, Edited by Ray d'Inverno.

\bibitem{Chop93c} M. W. Choptuik, 
``{\em Critical Behavior in Scalar Field Collapse},"
in {\it Deterministic Chaos in General Relativity}, Edited by D.
Hobill et al. (Plenum Press, New York, 1994), p. 155-175.

\bibitem{Golden0} G.I. Barenblatt, {\em Similarity, Self-Similarity, and
Intermediate Asymptotics} (Consultants Bureau, New York, 1979).

\bibitem{Golden1} N.  Goldenfeld, {\em Lectures on Phase Transitions and the
Renormalization Group} (Addison Wesley Publishing Company, New
York, 1992).

\bibitem{Gun00}  C. Gundlach,
``{\em Critical phenomena in gravitational collapse: Living
Reviews}," {\tt gr-qc/0001046} (2000), and references therein.

\bibitem{Wang01} A. Wang, ``{\em Critical Phenomena in Gravitational
Collapse: The Studies So Far}," {\tt gr-qc/0104073},  Braz. J.
Phys. {\bf 31}, 188 (2001), and references therein.

\bibitem{Cho03a} M.W. Choptuik, E.W. Hirschmann, S.L. Liebling, and F.
Pretorius, Phys. Rev. {\bf D68}, 044007 (2003).   

\bibitem{Cho03b} M.W. Choptuik, E.W. Hirschmann, S.L. Liebling, and F.
Pretorius,  Phys. Rev. Lett. {\bf 93}, 131101 (2004).
     
\bibitem{Wang03} A.Z. Wang,  Phys. Rev. {\bf D68}, 064006 (2003).

\bibitem{Even0} C. R. Evans and J. S. Coleman, Phys. Rev. Lett. {\bf 72},
1782 (1994).

\bibitem{Even1} T. Koike, T. Hara, and S. Adachi, Phys. Rev. Lett., {\bf
74}, 5170 (1995).

\bibitem{Even2} C. Gundlach, Phys. Rev. Lett. {\bf 75}, 3214
(1995).

\bibitem{Even3} E. W. Hirschmann and D. M. Eardley,  Phys. Rev. {\bf D52},
5850 (1995).

\bibitem{Brady02} P. R. Brady, M. W. Choptuik, C. Gundlach, and
D. W. Neilsen, Class. Quantum Grav. {\bf 19},  6359 (2002).

\bibitem{PC00}  F. Pretorius and M. W. Choptuik,  Phys. Rev. {\bf D62},
124012 (2000).

\bibitem{HO01} V. Husain and M. Olivier, Class. Quantum
Grav. {\bf 18}, L1 (2001).

\bibitem {Gar01} D. Garfinkle, Phys. Rev. {\bf D63}, 044007 (2001).

\bibitem{CF01a} G. Cl\'ement and A. Fabbri, Class. Quantum Grav. {\bf
18}, 3665 (2001). 

\bibitem{CF01b} G. Cl\'ement and A. Fabbri, Nucl. Phys. {\bf B630}, 269 (2002).

\bibitem{CF01c} M. Cavaglia, G. Clement, and A. Fabbri, Phys. Rev. {\bf D70}, 044010 (2004).

\bibitem{GG02} D. Garfinkle and C Gundlach, Phys. Rev. {\bf D66}, 044015
 (2002).

\bibitem{HWW04} E.W. Hirschmann, A.Z. Wang, and Y. Wu,
Class. Quant. Grav. {\bf 21}, 1791 (2004).

\bibitem {Chandra83}  S. Chandrasekhar, {\em The Mathematical Theory
of Black Holes} (Clarendon Press, Oxford University Press, Oxford,
1983).

\bibitem{CH89a} B. Carter and R.N. Henriksen, Ann.  Physique Suppl. {\bf 14},
47 (1989). 

\bibitem{CH89b} A.A. Coley, Class. Quantum Grav. {\bf 14}, 87 (1997).

\bibitem{CT71} M.E. Cahill and A.H. Taub, Commun. Math. Phys. {\bf 21}, 1 (1971).

\bibitem{Carlos02} C.F. Brandt, L.-M. Lin, J. F. Villas da Rocha and A.Z.
Wang, Inter. J. Mod. Phys. {\bf D11}, 155 (2002).
                                                                                
\bibitem{fluid0} B. J. Carr and A. A. Coley, Class. Quantum  Grav. {\bf 16},
R31 (1999). 

\bibitem{fluid1} H.-C. Kim, S.-H. Moon, and J. H. Yee, JHEP, {\bf 0202},
046 (2002).  

\bibitem{fluid2} H. Maeda, T. Harada, H. Iguchi, N. Okuyama, Prog. Theor. Phys.   {\bf 108},  819 (2002).

\bibitem{fluid3} H. Maeda, T. Harada, H. Iguchi, N. Okuyama, Prog. Theor. Phys.
 {\bf 110},  25 (2003).
                                                                                
\bibitem{Carlip} S. Carlip, {\em Quantum Gravity in $2+1$ Dimensions}
(Cambridge University Press, Cambridge, 1998).

\bibitem{Yasuda} A. Y. Miguelote, N. A. Tomimura, and A.Z. Wang, Gen. Rel.
Grav. {\bf 36}, 1883 (2004).

\bibitem{Pereira} F. I. M. Pereira, R. Chan , A. Z. Wang, Int. J. Mod. Phys. D,
{\bf 15}, 131 (2006).

\bibitem{CFJW05} R. Chan, M.F.A. da Silva, J.F. Villas da Rocha, A. Wang, 
Int. J. Mod. Phys. D, {\bf 14}, 1049 (2005).

\bibitem {Fatima0} M. Mars and J.M.M. Senovilla, Class. Quantum
Grav. {\bf 10}, 1633 (1993).

\bibitem{Fatima1} P.R.C.T. Pereira, N.O. Santos, and
A.Z. Wang, Class. Quantum Grav. {\bf 13}, 1641 (1996).

\bibitem{Fatima2} M.A.H. MacCallum
and N.O. Santos, Class. Quantum Grav. {\bf 15}, 1627 (1998).

\bibitem{Fatima3} M.A.H. MacCallum, Gen. Relativ. Grav. {\bf 30}, 131 (1998).

\bibitem{Fatima4} P.R.C.T.  Pereira, A.Z. Wang, Gen. Relativ. Grav. {\bf 32}, 
2189 (2000).

\bibitem{Fatima5} J. Carot, J.M.M. Senovilla, and R. Vera, Class. Quantum Grav. {\bf 16}, 3025 (1999).

\bibitem{Fatima6} A. Barnes, Class. Quantum Grav. {\bf 17}, 2605 (2000).

\bibitem{Fatima7} L.  Herrera, N.O. Santos, A.F.F. Teixeira, and A.Z. Wang, 
Class. Quantum Grav.  {\bf 18}, 3847 (2001).

\bibitem{Fatima8} A.Y. Miguelote, M.F.A. da Silva, A.Z. Wang,
and N.O. Santos, Class. Quantum Grav. {\bf 18}, 4569 (2001).

\bibitem{Fatima9} M.F.A. da Silva, L. Herrera, N.O. Santos, and A.Z. Wang, 
Class. Quantum Grav. {\bf 19}, 3809 (2002).

\bibitem{Kramer80}  D. Kramer, H. Stephani, E. Herlt, and M.
MacCallum, {\em Exact Solutions of Einstein's Field Equations}
(Cambridge University Press, Cambridge, England, 1980).

\bibitem{VS} A. Vilenkin and E. P. S. Shellard, {\it Cosmic String and Other
Topological Defects} (Cambridge University Press, Cambridge, 1994).

\end{thebibliography}
\end{document}